\documentclass[12pt, preprint]{aastex}
\usepackage{emulateapj5}
\usepackage{natbib}
\shortauthors{Strolger , Dahlen, \& Riess}
\shorttitle{Empirical Delay Time Distributions of Type Ia Supernovae}

\begin{document}
\title{Empirical Delay Time Distributions of Type Ia Supernovae\\ From The Extended GOODS/{\it HST} Supernova Survey}

\author{Louis-Gregory~Strolger,\footnote{Department of Physics and Astronomy, Western Kentucky University, 1906 College Heights Blvd., Bowling Green, KY 42101. {\tt louis.strolger@wku.edu}.}~~Tomas Dahlen,\footnote{Space Telescope Science Institute, 3700 San Martin Dr., Baltimore, MD 21218}~~and Adam G.~Riess\footnote{Department of Physics and Astronomy, Johns Hopkins University, 3400 North Charles St., Baltimore, MD~~21218.}}

\begin{abstract}
Using the {\it Hubble Space Telescope} ACS imaging of the GOODS North and South fields during Cycles 11, 12, and 13, we derive empirical constraints on the delay-time distribution function for type Ia supernovae. We extend our previous analysis to the three-year sample of 56 SNe Ia over the range $0.2<z<1.8$, using a Markov chain Monte Carlo to determine the best-fit unimodal delay-time distribution function. The test, which ultimately compares the star formation rate density history to the unbinned volumetric SN~Ia rate history from the GOODS/HST-SN survey, reveals a SN~Ia delay-time distribution that is tightly confined to $3-4$ Gyrs (to $>95\%$ confidence). This result is difficult to resolve with any intrinsic delay-time distribution function (bimodal or otherwise), in which a substantial fraction (e.g., $>10\%$) of events are ``prompt'', requiring less than approximately 1 Gyr to develop from formation to explosion. The result is, however, strongly motivated by the decline in the number of SNe~Ia at $z>1.2$.  Sub-samples of the HST-SN data confined to lower redshifts ($z<1$) show plausible delay-time distributions that are dominated by prompt events, which is more consistent with results from low-redshift supernova samples and supernova host galaxy properties. Scenarios in which a substantial fraction of $z>1.2$ supernovae are extraordinarily obscured by dust may partly explain the differences in low-$z$ and high-$z$ results. Other possible resolutions may include environmental dependencies (such as gas-phase metallicity) that affect the progenitor mechanism efficiency, especially in the early universe.

\end{abstract}

\keywords{supernovae: general-- {\sc Accepted to the Astrophysical Journal}}

\section{Introduction}

The progenitor systems and mechanisms responsible for type Ia supernovae (SNe~Ia) remains unresolved. The general consensus is that SNe~Ia stem from C+O white dwarf stars (WD) that accrete mass until they exceed the the electron degeneracy pressure limit in their cores, marked by the Chandrasekhar mass limit of approximately $1.4\,M_{\odot}$. But the details of how the additional mass is accumulated, or specifically what the donor mass source is, remains largely ambiguous. Very broadly, progenitor models are categorized as either involving mass accretion from companion stars (typically red-giant stars) in single degenerate (SD) scenarios, or pairs of WDs that merge through coalescence or collisions in double degenerate (DD) scenarios (see~\citealt{2000ApJ...528..108Y} for a review). The various detailed modeling of these scenarios have thus far provided adequate agreement with the observed spectra and light-curves of SN~Ia events, largely due to gross requirement radioactive Fe-peak material to power the event (see \citealt{2009arXiv0907.3196R} and \citealt{2009arXiv0907.3915R} for recent examples involving collisional WD mergers). There remains much uncertainty as to which scenarios are actually employed to make SNe~Ia.

Unlike the progenitors of core-collapse supernovae that are now directly found through deep archival imaging at a rate of a few per year, SNe~Ia are much rarer (by about a factor of ten in typical low-$z$ galaxies) and their progenitors are much fainter (by a factor of several million), making it extraordinarily unlikely that these progenitors will be similarly resolved in the near future.  Nonetheless, meaningful constraints on SN~Ia progenitors can be drawn from the resolvable hosts environments of these events. Age limits on the stellar population, the rate of formation of new stars, and the range of chemical enrichment in the environment of the event all provide implied but important constraints  on the nature of progenitor systems--- the types of stars involved, and the physical mechanisms employed to result in these luminous explosions. It is therefore expected that correlations drawn from these environmental characteristics and characteristics of SNe~Ia (e.g., event luminosity, or event production) will eventually illuminate how SNe~Ia are formed. 

However presently, the analysis of SN~Ia rates in high and low redshift galaxies show inconsistent results on the implied progenitor mechanisms responsible for producing these important cosmological tools.  At its heart, the discrepancy hinges on two important factors: (1) the incubation time of SNe~Ia (commonly called the ``delay time''), or the time required for a progenitor system to develop into an explosion from a single episode of star-formation; and (2) the metallicity of the progenitor at formation, and its impact on either production efficiency or event luminosity. While these are conceptually measurable factors in low-redshift ($z<0.1$) galaxies, attempts to do so (e.g.,~\citealt{2008ApJ...685..752G, 2009ApJ...691..661H}) have been muddled by two degenerate effects: (1) population age, which steadily increases the range of metallicity within a given environment, and (2) rate of active star formation, which mix-up the time between events and progenitor formation. 

It is expected that at  high redshifts ($z > 1$) these confusion effects should become less severe as the age of the parent population is limited by the age of the universe at $z>1$ to be less than 6 Gyr old. Moreover, any substantial bulk delay-time would limit the SN~Ia  rate in the highest redshift regimes as the Universe would not be old enough to produce them. Thus observations in the highest redshifts could provide the best leverage in discerning the ages of SN~Ia progenitors. However the implied  delay-time distribution from an investigation of the first year (Cycle 11) imaging of the GOODS North and South fields with the {\it Hubble Space Telescope} (HST) and ACS ~\citep{2004ApJ...613..200S} were on average very long (3 to 4 Gyr), and inconsistent with the relatively short times predicted from binary star evolutionary models (cf.~\citealt{2008IAUS..252..349H}). They were also  difficult to reconcile with several observations of SNe~Ia in low-$z$ galaxies (cf.~\citealt{2000AJ....120.1479H,2005A&A...433..807M, 2005ApJ...629L..85S, 2008ApJ...685..752G}) that suggest there are at least two mechanisms for SN~Ia production, one which could require a few Gyr incubation, and a second which is much prompter, requiring only a few 100 Myr of incubation. Additionally there was seemingly little support for the implied trend of an increasingly prompt-dominated mechanism as observations are pushed to higher redshifts from the ground~\citep{2007ApJ...667L..37H}.

With just 25 events from the Cycle 11 data, and few events at $z>1.2$, it is possible the sample provided too little statistical certainty to ascertain the inherent delay time distribution function. In addition, the tested models were simplistic and did little to address more than the average incubation times for SN~Ia progenitors. More flexible functionality to the model tests would do better to assess the delay time distribution. 

The  addition of the Cycle 12 \& 13 HST-SN data provides an opportunity to revisit the best delay-time model analysis on a larger statistical sample with a more complete analysis. Here we present the extension of the investigation of~\citet{2004ApJ...613..200S}, applied to the complete three-year (Cycles 11, 12, and 13) sample of 56 SNe~Ia. We compare the observed redshifts of the SNe~Ia  (in the range $0.2 < z < 1.8$) to event redshift distributions forecasted from model delay-time distribution functions, assuming the star formation rate density model used in~\citet{2004ApJ...613..200S}. Through a Markov chain Monte Carlo test, we determine the most likely model delay-time function, empirically implying the bulk distribution of incubation times of SNe~Ia. In \S\ref{hzrates} we discuss measurements of the SN~Ia rate history, and its relation to the rate of star formation and the delay times of SNe~Ia. In \S\ref{model} we describe the model delay-time tests. In \S\ref{results} we show the highest likelihood delay-time distribution model from the HST-SN data. And in \S\ref{discussion} we discuss the results in comparison to other determinations of the ages of SN~Ia progenitor systems.

\section{The SN~Ia Rate History}\label{hzrates}
The various SD and DD scenarios are expected to involve very different characteristic development times and distributions (\citealt{2000ApJ...528..108Y,2005A&A...441.1055G}). The rate of SNe~Ia at any given epoch should therefore mimic the stellar birthrate at earlier cosmic epochs, shifted and convolved with the delay-time distribution function, $\Phi(\tau)$, that is set by the progenitor mechanism scenario. Here $\tau$ is the elapsed time between the progenitor system formation and explosion as a SN Ia event. The volumetric SN~Ia rate (R$_{\rm Ia}$, in units of events yr$^{-1}$ Mpc$^{-3}$ $h^3$) at any specific cosmic age can therefore be described by, 
\begin{equation}
{\rm R}_{\rm Ia}(t)\,=\,\int_{t_0}^t \Phi(t-t')\,{\rm \dot{\rho}_{\star}}(t')\biggl[ {\rm A}_{\rm Ia}\int_M \xi(M)\,dM\biggr]\,dt', 
\label{eqn:snriadef}
\end{equation}
\noindent where the stellar birth rate is the combination of the star formation rate density [hereafter $\dot{\rho}_{\star}(t)$, in M$_{\odot}$ yr$^{-1}$ Mpc$^{-3}$ $h$], and the initial mass function [hereafter $\xi(M)$, in M$_{\odot}^{-1}$].  This equation is easily translated to redshift space by defining $t$ as the age of the universe at redshift $z$, and setting $t_0$ as the age of the universe when the first stars were born, corresponding to $z\approx10$. It follows that $\int\dot{\rho}_{\star}(t)\,dt\equiv\int\dot{\rho}_{\star}(z)\, dz$.  The $\Phi(\tau)$ distribution function is normalized so that the integral sum of the function over all incubation times is unity.

Although there is evidence that suggests $\xi(M)$ may vary with cosmic time~\citep{2008MNRAS.385..147D,2008ApJ...674...29V}, these variations are notably small (with small changes in power-law slope) in the $\lesssim8~\rm{M}_{\odot}$ mass range for SN~Ia progenitor system stars. For the purposes of this investigation, $\xi(M)$ is assumed to be more or less invariant with cosmic time. It is also assumed that the SN~Ia mechanism, although not ubiquitously affecting all stars in the progenitor mass range,  affects essentially the same fraction of stars in $\xi(M)$ in all cosmic epochs, thus allowing A$_{\rm Ia}(t)=$ constant. With these preliminary assumptions, we define:
\begin{equation}
	\varepsilon\equiv\biggl[{\rm A}_{\rm Ia}\int_M \xi(M)\,dM\biggr],
\end{equation}
\noindent where $\varepsilon$ is a constant that describes the number of SNe~Ia produced per M$_{\odot}$ formed, or an efficiency for the stellar progenitor population in actually producing SN~Ia events, as presumably not all WDs become SNe~Ia. The rate of SNe~Ia can therefore be simplified as:
\begin{equation}
{\rm R}_{\rm Ia}(t)=\varepsilon\,\int_{t_0}^t \Phi(t-t')\,{\rm \dot{\rho}_{\star}}(t')\,dt',
\label{ria}
\end{equation}
\noindent where the shape of the rate distribution with time is only dependent on star formation rate density history and the distribution of delay times.

\begin{figure*}[t] 
   \centering
   \includegraphics[width=6.0in]{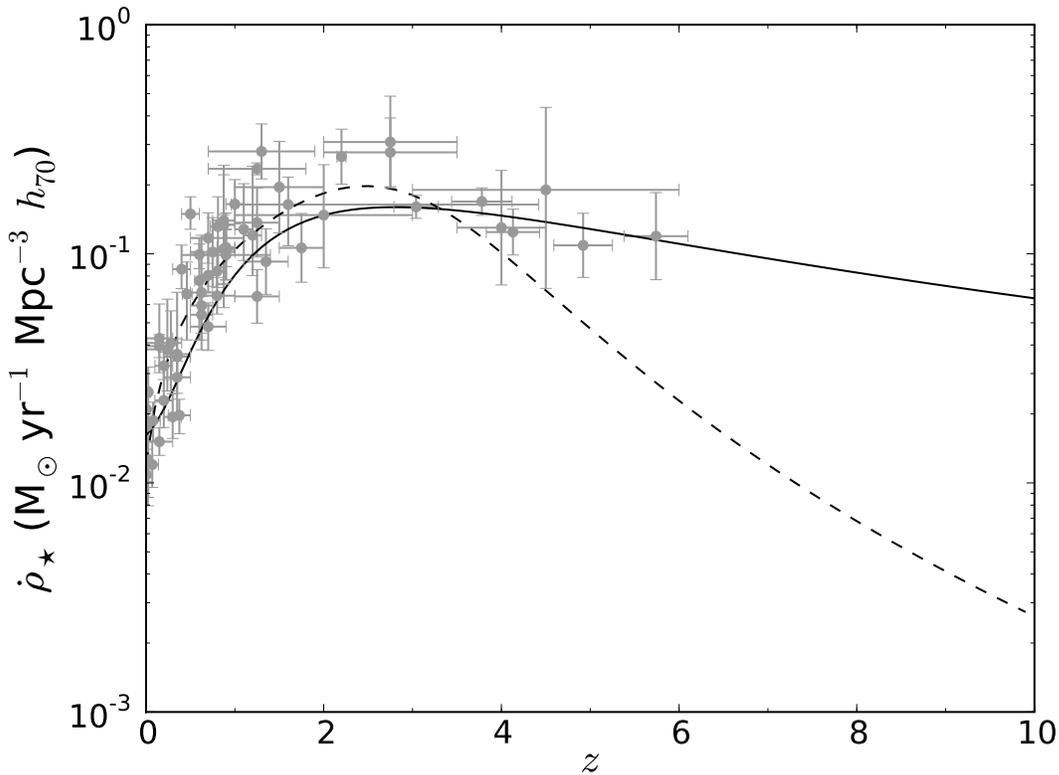}
   \caption{\footnotesize The semi-analytical $\dot{\rho}_{\star}(z)$ model is shown (solid line), representing the compilation of measurements (grey points) from \citealt{2004ApJ...615..209H}). For comparison, the best-fitting the \citet{2001MNRAS.326..255C} parametric function (dashed line) is shown. This corrects an inaccurate representation shown in~\citet{2009NewA...14..638V}. Recent constrains from the HST+WFC3/IR Early Release data~\citep{2010ApJ...709L.133B,2010ApJ...709L..16O} are not included in the model fit.}\label{fig:sfr}
\end{figure*}

\begin{figure*}[t] 
   \centering
   \includegraphics[width=6.0in]{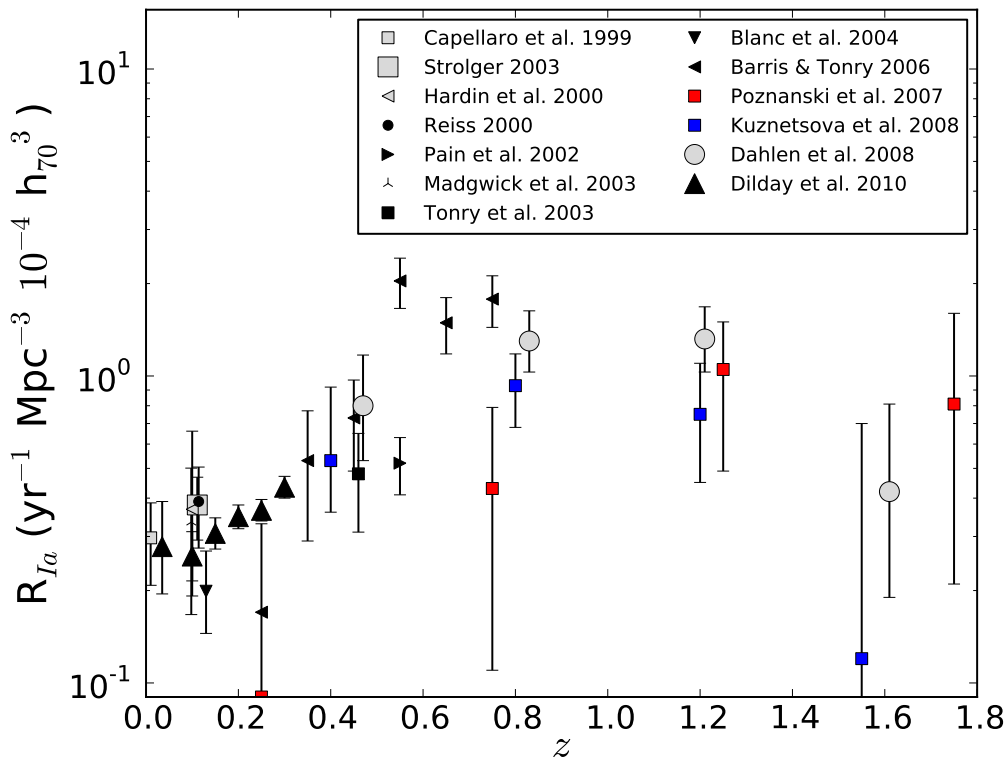}
   \caption{\footnotesize Various type Ia supernova rate measures as a function of redshift. Measures are averages over small redshift bins centered at the measured point. The widths of these bins are not shown for clarity in the diagram.}\label{d08}
   \end{figure*}

With the advancing knowledge of $\dot{\rho}_{\star}(z)$ and R$_{\rm Ia}(z)$, it is plausible to attempt to ``deconvolve'' Equation~\ref{ria}  to learn about $\Phi(\tau)$, and thus constrain the nature of SN~Ia progenitor systems, which is the goal of this investigation. To this end, we are fortunate that the star formation rate density history has been mostly resolved to at least $z<6$, with the compilation data presented in \citet{2004ApJ...615..209H} and \citet{2006ApJ...651..142H} (see Figure~\ref{fig:sfr} of this paper). Although unfortunately, the volumetric SN~Ia rate history is much farther from consensus. While  it is expected that the SN~Ia rate should demonstrate some increase with lookback time (e.g., \citealt{1997ApJ...486..110J}), the scattered and often disagreeing rate measurements to date (most shown in Figure~\ref{d08}) has made this evolution difficult to resolve.~\footnote{However, recent results from the Sloan Digital Sky Survey-II~\citep{2010arXiv1001.4995D} show the promise of the large-scale surveys currently underway and in the near future.} Figure~\ref{d08} shows rate measurements from various authors in several redshift regimes~\citep{1999A&A...351..459C, Reiss2000,2000A&A...362..419H,2002ApJ...577..120P,2003PhDT........14S,2003ApJ...594....1T,2003ApJ...599L..33M,2004A&A...423..881B,2006ApJ...637..427B,2007MNRAS.382.1169P,2008ApJ...673..981K,2008ApJ...681..462D,2010arXiv1001.4995D}. It is presently unclear where the point-to-point variation in the rate measures stem from, although the most likely culprits are survey completenesses (i.e., {\it were all discoverable SNe~Ia identified?}), and differences in applied corrections for undiscovered events, typically referred to as control-time corrections.

These discrepancies in SN~Ia rate measures pose the greatest limitation in determining $\Phi(\tau)$ empirically. The HST-SN survey rates~\citep{2004ApJ...613..189D, 2008ApJ...681..462D} have at least given a self-consistent measure of the SN~Ia rates over a wide redshift range of $0.2<z<1.8$ (see Figure~\ref{d08}), making a determination for $\Phi(\tau)$ from these rate measures alone reasonable. Additionally, the star formation rate density compilation data in the $z>3$ range is largely from the same GOODS/UDF fields~\citep{2004ApJ...600L..93G, 2007ApJ...670..928B,2009ApJ...705..936B}, making the GOODS/HST-SN dataset most ideal for empirically probing the intrinsic $\Phi(\tau)$ function of SNe~Ia, relatively free of the point-to-point R$_{\rm Ia}(z)$ biases, and biases due to cosmic variance.

\section{The Model Tests}\label{model}\label{results}
While the goal of this investigation is to determine the $\Phi(\tau)$ function from the~\citet{2008ApJ...681..462D} data, a direct comparison to just the binned rate measurements is unsuitable as it provides only one or two datapoints to assess the production of SNe~Ia at $z>1.2$. A more robust comparison can be made to the individual SN~Ia events that went in to the volumetric rate calculations. In general, the volumetric SN~Ia rate is determined by comparing the survey SN~Ia yield to the product of volume and the effective period (or control time) of the survey.  In a redshift range ($\Delta z=z_2-z_1$) centered at a given redshift ($z$), the rate is generally expressed by:
\begin{equation}
	{\rm R}_{\rm Ia}(z)=\frac{\sum_{i=z_1}^{z_2} {\rm N}_{\rm Ia}(z_i)}{[\sum_{i=z_1}^{z_2} t'_c(z_i)]\,\Delta V(z)},
	\label{ria2}
\end{equation}
\noindent where the N$_{\rm Ia}(z)$ is the yield in the $z-$bin, $t'_c(z)$ is the control time, corrected for time dilation, and $\Delta V(z)$ is the conically sliced volume from $z_2$ to $z_1$ (the bin size). A complete description of these parameters and how they are determined for the HST-SN survey, including $t'_c(z)$, is presented in~\citet{2004ApJ...613..200S}.  

We probe for an empirical $\Phi(\tau)$ function in a method also described more completely in~\citet{2004ApJ...613..200S}. To summarize, with assumed $\dot{\rho}_{\star}(z)$ and $\Phi(\tau)$ models, we use Equations~\ref{ria} and~\ref{ria2} to predict the expected redshift distribution of SNe~Ia for the survey. This result is compared to the observed redshifts of each SN~Ia to produce a conditional probability test in an application of Bayes' method, where:
\begin{eqnarray}
	\rm{P} [\dot{\rho}_{\star}(z), \Phi(\tau)\vert\rm{Data}]&\approx&\rm{P}[\rm{Data}\vert \dot{\rho}_{\star}(z),\Phi(\tau)],\nonumber\\
	&=&\prod_{i=1}^{56}{\rm N}_{\rm Ia}(z_{i}),\nonumber\\
	&=&\prod_{i=1}^{56}{\rm R}_{\rm Ia}(z_{i})\,t'_{c}(z_{i})\,\Delta V(z_i).\label{eqn:probability}
\end{eqnarray}
The predicted number distribution, given the assumptions on the models, then serves as a probability function for finding SNe~Ia at the specific redshifts in which we have found them. By this method, we maximize our leverage on the best $\Phi(\tau)$ model by using all 56 SNe~Ia, rather than just the four binned rate measurements. 

We normalize the probability distributions to serve as a relative likelihood statistic. Changes in the input model parameters will allow changes in the likelihood with redshift. For this investigation, we use the extinction-corrected $\dot{\rho}_{\star}(z)$ determined from a semi-analytical fit to various measures (functional form shown in~\citealt{2004ApJ...613..200S}). The correction to extinction internal to each galaxy is consistently  applied to both the published star formation rate densities and to the control times for the SN~Ia rate calculations, both in \citet{2008ApJ...681..462D} and in this analysis. It is assumed that the $\dot{\rho}_{\star}(z)$ model and all other dependencies (e.g., $\Omega_{M}$, $\Omega_{\Lambda}$, $H_0$, and survey parameters) are sufficiently well determined that their uncertainties do not significantly contribute to the overall probability.

The test method is to select a model $\Phi(\tau)$, calculate the N$_{\rm Ia}(z)$, and determine a relative Bayesian likelihood of the chosen model from a comparison to the observed redshifts of our discovered SNe~Ia. The likelihood values maximize when the predicted redshift distribution matches the observed redshifts, and minimize (or become zero) if for example a discovered SN~Ia's redshift falls outside the predicted distribution. This high-risk test gives significant leverage in selecting the most likely $\Phi(\tau)$ when dealing with a statistically limited sample, free from biases associated with binning data for $\chi^2$ or K-S tests on observed redshift distributions or rate measurements. Unfortunately, it also has a limited ability to address which $\Phi(\tau)$ models are implausible given the data, and therefore such interpretations should be made cautiously.

In~\citet{2004ApJ...613..200S} we examined a few simple single-parameter models (Gaussian models with variance tied to the mean, and exponential models), linearly iterating through the characteristic delay time, $\bar{\tau}$ (describing the Gaussian mean time or the e-fold time), for a maximum likelihood. In that preliminary Bayesian test, the HST-SN data provided the highest likelihoods for Gaussian $\Phi(\tau)$ models with a mean of $\bar{\tau}\approx3.5$ Gyr and $\sigma_\tau=0.2$, seemingly indicating a single, highly delayed mechanism  for all SN~Ia production.  Models with incubation times less than 2.0 Gyr where inconsistent with the data to 95\% confidence. However, the preliminary investigation was limited in that the tests could not assess large asymmetric skews (other than with an exponential decay), or the possibility multi-modal populations. Most of the power of the previous tests was in determining the likeliest mean of incubation times, and there was limited ability to reveal a more complex $\Phi(\tau)$ distribution.

\begin{figure*}[t] 
   \centering
   \includegraphics[width=6.0in]{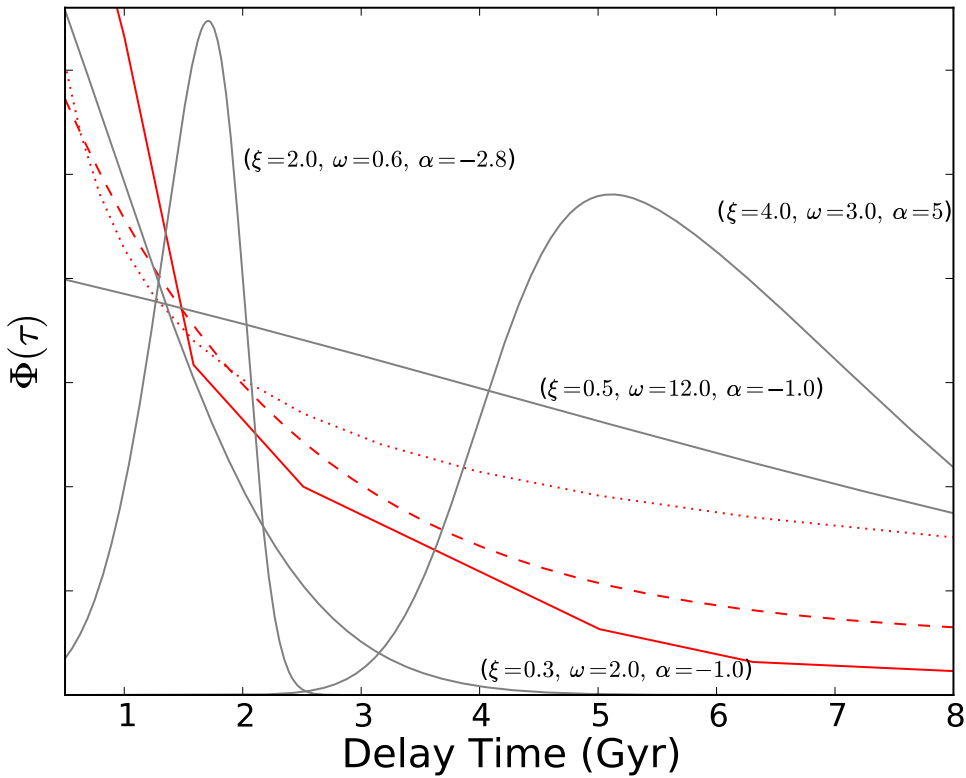}
   \caption{\footnotesize An illustration of the flexibility of the skew-normal model used in this analysis (from Equation~\ref{eqn:model}). The grey solid lines exemplify $\Phi(\tau)$ models for different choices of $\xi$, $\omega$, and $\alpha$, while the red lines are example model $\Phi(\tau)$ from the literature (\citealt{2008MNRAS.388..829G}, solid;~\citealt{2008ApJ...683L..25P}, dotted; and~\citealt{2005ApJ...629L..85S}, dashed).}\label{dtdex}
\end{figure*}

As the natural next step in this investigation, we now test a more robust delay-time model, capable of more accurately reproducing the theoretical distributions for SD and DD models at one extreme, and $\delta$-function delay times at the other. The unimodal, skew-normal $\Phi(\tau)$ function is defined as:
\begin{equation}
	\Phi(\tau)=\frac{1}{\omega\pi}\,\exp\biggl(\frac{-(\tau-\xi)^2}{2\omega^2}\biggr)\int_{-\infty}^{\alpha (\frac{\tau-\xi}{\omega})} \exp\biggl(\frac{-t'^2}{2}\biggr)\,dt',
\label{eqn:model}
\end{equation}
where location ($\xi$),\footnote{Different from the initial mass function, $\xi(M)$.} scale ($\omega^2$), and shape ($\alpha$) define the mode time ($\bar{\tau}$, as defined in the previous tests), variance ($\sigma^2$), skewness ($\gamma_1$), and kurtosis ($\gamma_2$) of the model function by,
\begin{eqnarray*}
 \bar{\tau}&=&\xi+\omega\delta\sqrt{\frac{2}{\pi}},\\
 \delta&=&\frac{\alpha}{\sqrt{1+\alpha^2}},\\
 \sigma^2&=&\omega^2\biggl(1-\frac{2\delta^2}{\pi}\biggr),\\
\gamma_1&=&\frac{1}{2}(4-\pi)\frac{(\delta\sqrt{2/\pi})^3}{(1-2\delta^2/\pi)^{3/2}},\\
\gamma_2&=&2(\pi-3)\frac{(\delta\sqrt{2/\pi})^4}{(1-2\delta^2/\pi)^{2}}.\\
\end{eqnarray*}
\noindent An illustration of the diversity of testable delay time distribution functions from Equation~\ref{eqn:model} can be seen in Figure~\ref{dtdex}, where the four $\Phi(\tau)$ models shown are created from independent choices of  $\xi$, $\omega$, and $\alpha$ model parameters. 

To test which model parameters values best fit our data, $\xi$, $\omega$, and $\alpha$ have been jointly explored in a Metropolis-Hastings Markov chain Monte Carlo~\citep[hereafter MCMC]{Hay70,Met53} to find the most likely regions of the 3-D parameter space. To illustrate the MCMC test--- a three-dimensional array was constructed for all $\xi$ (in range $-10\dots+10$), $\omega$ (range $0\dots+10$), and $\alpha$ ($-10\dots+10$), in intervals of $0.1$. The MCMC values in this array were all initially set to zero. An initial starting point in $\xi$, $\omega$, and $\alpha$ was randomly selected, and a relative Bayesian likelihood value was determined from Equation~\ref{eqn:probability} at this starting point. The algorithm then randomly determined a step in all parameters simultaneously, where each step size (and direction) was determined from a normal distribution, centered at zero with $\sigma=0.25$. In this way, 68\% of the steps were $\Delta{(\xi,\omega,\alpha)}\le\pm0.25$ (rounded to the 0.1 interval in the grid). The conditional test of Equation~\ref{eqn:probability} was then run on the new $\xi+\Delta\xi$, $\omega+\Delta\omega$, and $\alpha+\Delta\alpha$ position. If the Bayesian likelihood value was greater at the new position than it was for the starting position (or equal to it), the algorithm incremented the MCMC array value in the new position by 1, adopted the new position as the starting point, and repeated the process of selecting a new step for evaluation. If, however, the Bayesian likelihood for the new position was less than the value at the starting position, then the MCMC at the starting position was instead incremented by the ratio of likelihood values.\footnote{The ratio is the Bayesian likelihood value (Equation~\ref{eqn:probability}) at the new position over the likelihood at the starting position.} The starting position remained unchanged, and a new step was randomly selected and evaluated. 

\begin{figure*}[t]
   	\centering
   	\includegraphics[width=6in]{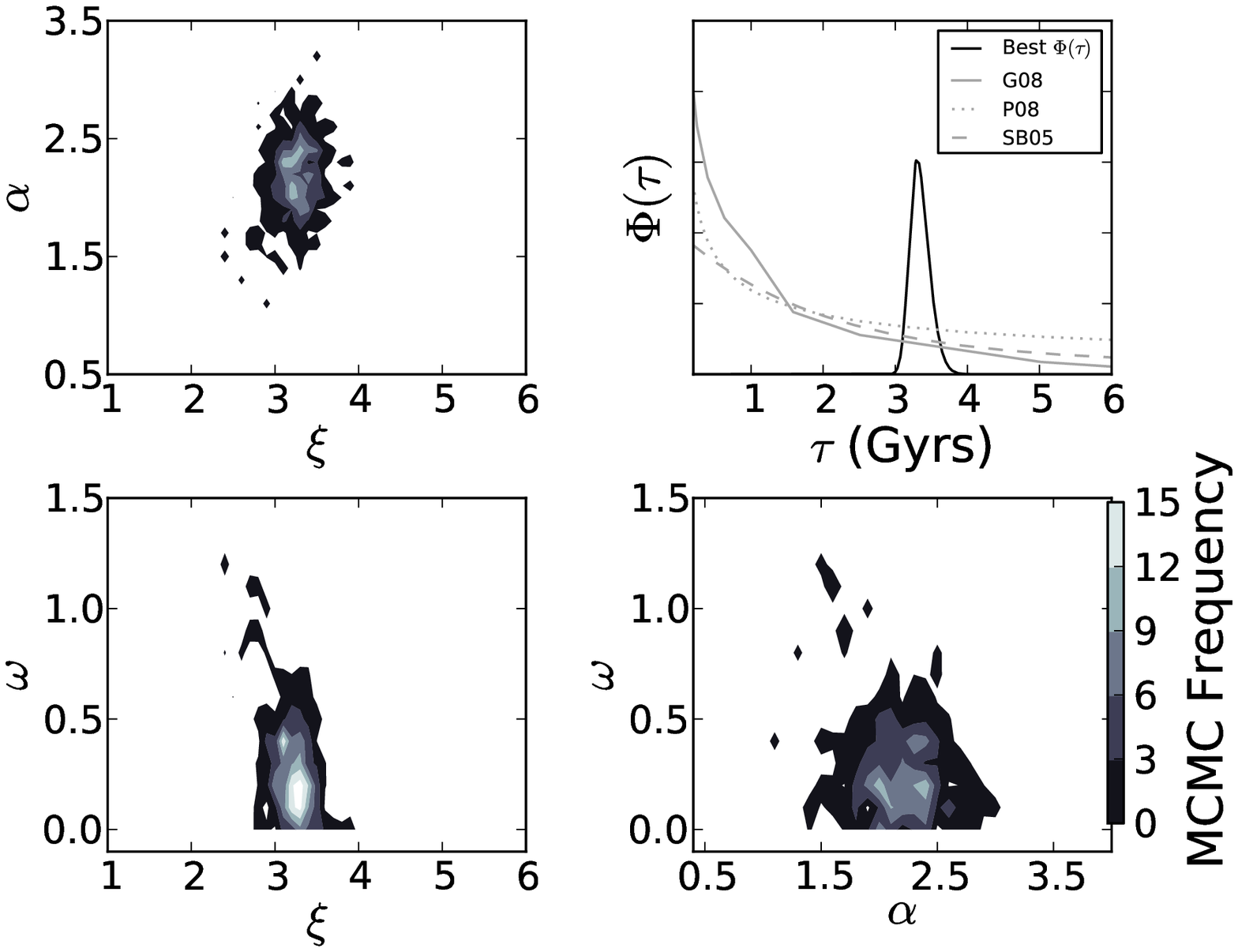}
\caption{\footnotesize Results of the Markov chain Monte Carlo test of three-parameter skew-normal distribution, where $\xi$, $\omega$, and $\alpha$ (which describe the shape of the delay time function) are explored. The best-fit $\Phi(\tau)$ has been determined from centroid of the 95\% likelihood region, and is shown in in the upper-right panel (solid line). Shown for comparison is the~\citet{2008MNRAS.388..829G} single degenerate $\Phi(\tau)$ model (G08; solid grey line), the~\citet{2005ApJ...629L..85S} bimodal model (SB05; dashed grey line), and the~\citet{2008ApJ...683L..25P} white-dwarf availability model (P08; dotted grey line). The scalings of the $\Phi(\tau)$ models are arbitrary.}\label{contour1}
\end{figure*}

\begin{figure*}[t] 
   \centering
   \includegraphics[width=6.0in]{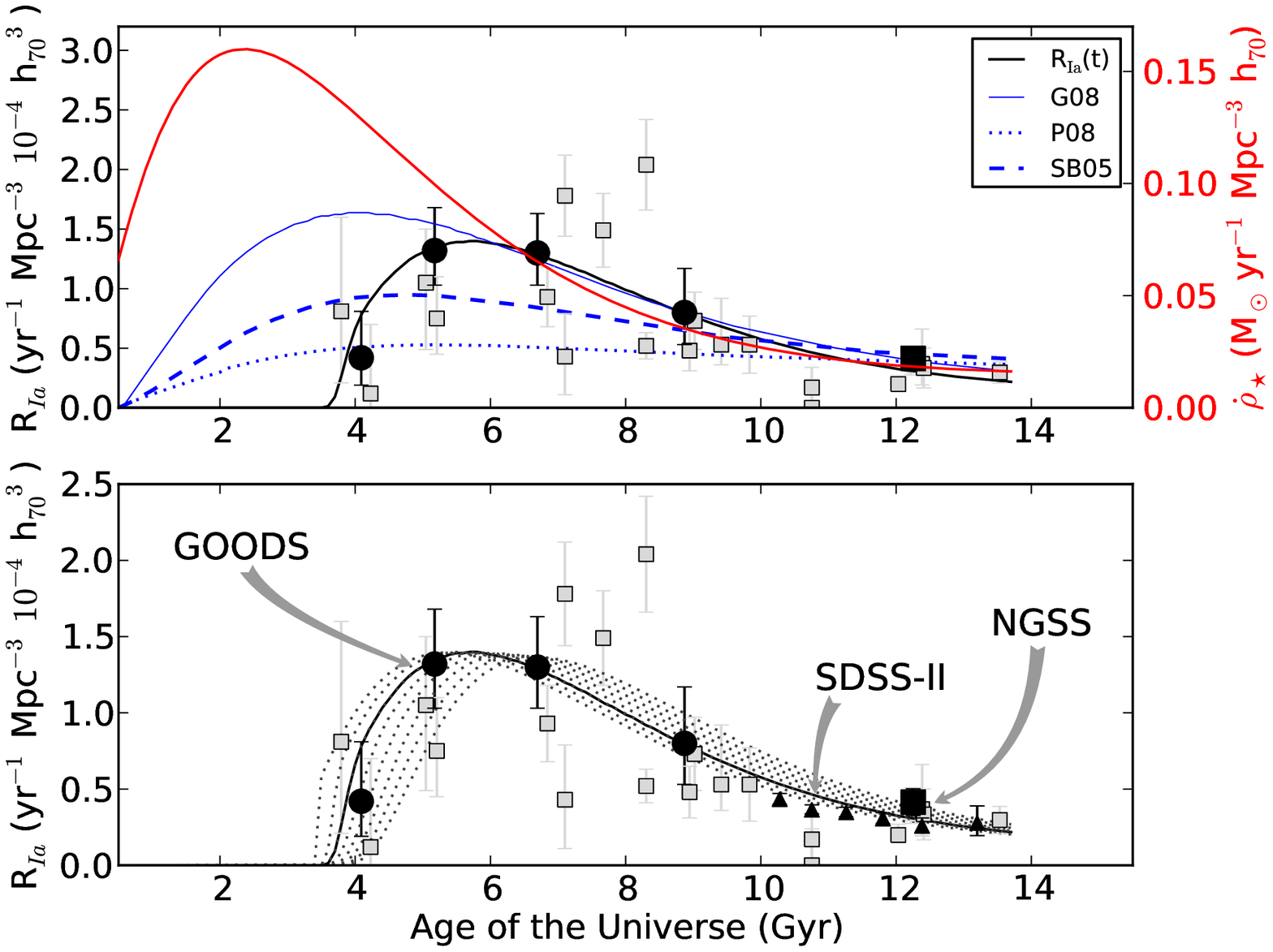}
   \caption{\footnotesize Model type Ia supernova rates compared to rate measurements, displayed in terms of the age of the universe (concordance model cosmology is assumed). {\it Upper panel:}~The solid black line is the best-fitting R$_{Ia}(t)$ from the $\Phi(\tau)$ model derived in this investigation, with an effective mean $\tau=3.5$ Gyr. For comparison, the blue lines show model R$_{Ia}(t)$ using $\Phi(\tau)$ models from~\citet[solid]{2008MNRAS.388..829G},~\citet[dotted]{2008ApJ...683L..25P}, and~\citet[dashed]{2005ApJ...629L..85S}. The red line (and secondary axis) is the model $\dot{\rho}_{\star}(t)$. {\it Lower panel:}~ The best-fit model shown with a confidence region (dotted black lines).}\label{d08a}
\end{figure*}

After a ``burn in'' of 125 iterations (which were discarded), over 1000 test iterations were made where Equation~\ref{eqn:probability} served as the conditional likelihood between each step in the process. The MCMC array values thus built up at points in the grid with highest likelihood. Figure~\ref{contour1} shows the most likely region of the MCMC test, in which $>95\%$ of MCMC values lie.  The peak of the confidence region is located at $\xi=3.2^{+0.8}_{-0.4}$, $\omega=0.2^{+0.8}_{-0.2}$, and $\alpha=2.2\pm1.4$ where the uncertainties are an approximation of the 95\% confidence region. These model parameters represent a $\Phi(\tau)$ with mode $\bar{\tau}=3.4$ Gyr and $\sigma=0.14$, shown in the upper right of Figure~\ref{contour1}. The skew ($\gamma_1=0.51$), and kurtosis ($\gamma_2=0.35$), are nearly irrelevant for such a narrow distribution. The test indicates a $\Phi(\tau)$ that is surprisingly consistent with the best-fit narrow-gaussian model of \citet{2004ApJ...613..200S} in that the mean incubation time is approximately 3.5 Gyr, and the entire distribution in contained within a 3 to 4 Gyr span. The implication is, once again,  that there is only a single mechanism for SN~Ia production that requires {\it almost exactly} 3.5 Gyr to develop WD progenitors into SNe~Ia. The lack of width in the $\Phi(\tau)$ or a strong asymmetry is incorroborative of multiple components in this high redshift data. The SN~Ia rate history, calculated from the MCMC best-fit delay-time distribution, is shown in Figure~\ref{d08a}. The comparison rate models shown in the figure are scaled to match the observed SN Ia rates at $z\la0.1$. 

\section{Discussion}~\label{discussion}
The strong preference for a very specific delay time (with little variation) is uncanny, given the wide freedom of the tested delay-time distribution model. If $\Phi(\tau)$ were intrinsically bimodal (as in \citealt{2006MNRAS.370..773M} or~\citealt{2005ApJ...629L..85S}, for example), the test should have preferred a substantially wider and greatly skewed $\Phi(\xi,\omega,\alpha)$ model, similar to what is shown in Figure~\ref{dtdex}. A similarly exponential $\Phi(\tau)$ would have been expected if the SN~Ia rate were directly tied to just the availability of WD at any given epoch, as is suggested by the $\log[\Phi(\tau)]\approx-0.5\,\log(\tau)+{\rm constant}$ model derived from a calculation of WD formation rates, and dominated by the main-sequence lifetimes of their $3-8$ M$_{\odot}$ progenitor stars~\citep{2008ApJ...683L..25P}. But again, the MCMC test here does not find this to be the preferred model. The implication is that SN~Ia are predominately characterized by a single channel (or mechanism) for explosion that requires almost exactly the same incubation time from formation to explosion.

This interpretation, however,  is difficult to reconcile with various results at lower redshifts. Studies in $z<0.1$ galaxies have shown strong evidence for SN~Ia heterogeneity that extends beyond the luminosity dispersion characterized by the luminosity-lightcurve width relations~\citep{1993ApJ...413L.105P,1999AJ....118.1766P}, including low-luminosity SNe~Ia with strong Ti absorption (e.g., SN~1991bg,   \citealt{1993Natur.365..728R}), high-luminosity SNe~Ia with weak Si II absorption at early times (e.g., SN~1991T, \citealt{1992ApJ...384L..15F}), and several examples of individually peculiar supernovae. In addition, rate measures in low-$z$ galaxies show that late-type galaxies are rather prolific producers of SNe~Ia, producing an order of magnitude more events than early-type galaxies of the same total stellar mass~\citep{2005A&A...433..807M}. Moreover, trends show that the most luminous SNe~Ia are largely absent in early-type galaxies, and low-luminosity events are deficient in late-type galaxies~\citep{2000AJ....120.1479H, 2004MNRAS.349.1344A}. The implication has been that the rate of recent star-formation has a significant impact on the production of SNe~Ia, and that the overall rate of SNe~Ia is a sum of prompt and delayed components.~\citet{2005ApJ...629L..85S} have argued that a two-component model with a large contribution  of prompt SNe~Ia ($20-40\%$ of all SNe~Ia) can resolve the inter-cluster Fe content and [O/Fe] abundance in the Milky Way and remain consistent with the observed low-$z$ rates. Moreover, the rate at higher redshifts ($\bar{z} \sim 0.6$) from the  Supernova Legacy Survey also support the two-component model, but with a prompt component increased in its fraction to as high as $80-90\%$  by $z\la1$~\citep{2006AJ....132.1126N}.

Indeed, a more direct relationship between SN~Ia rates and star formation rates (with little-to-no delay) appears possible given notable similarity between the $\dot{\rho}_{\star}(t)$ function and most R$_{\rm Ia}(t)$ measurements  at $z<1$ in Figure~\ref{d08a}, using a scaling where approximately one SN~Ia is produced for every 500 M$_{\odot}$ created with essentially zero delay.\footnote{This is effectively reduces Equation~\ref{ria} to R$_{\rm Ia}(t)=\varepsilon\,\dot{\rho}_{\star}(t)$, where $\varepsilon=0.002\,\rm{M}_{\odot}^{-1}$.} It is, therefore, interesting to address what results when the HST-SN sample is cut to $z<1.0$ and our MCMC test are re-done.

\subsection{Cuts on the Data: The $z<1$ Sample}
\begin{figure*}[t]
   	\centering
   	\includegraphics[width=6in]{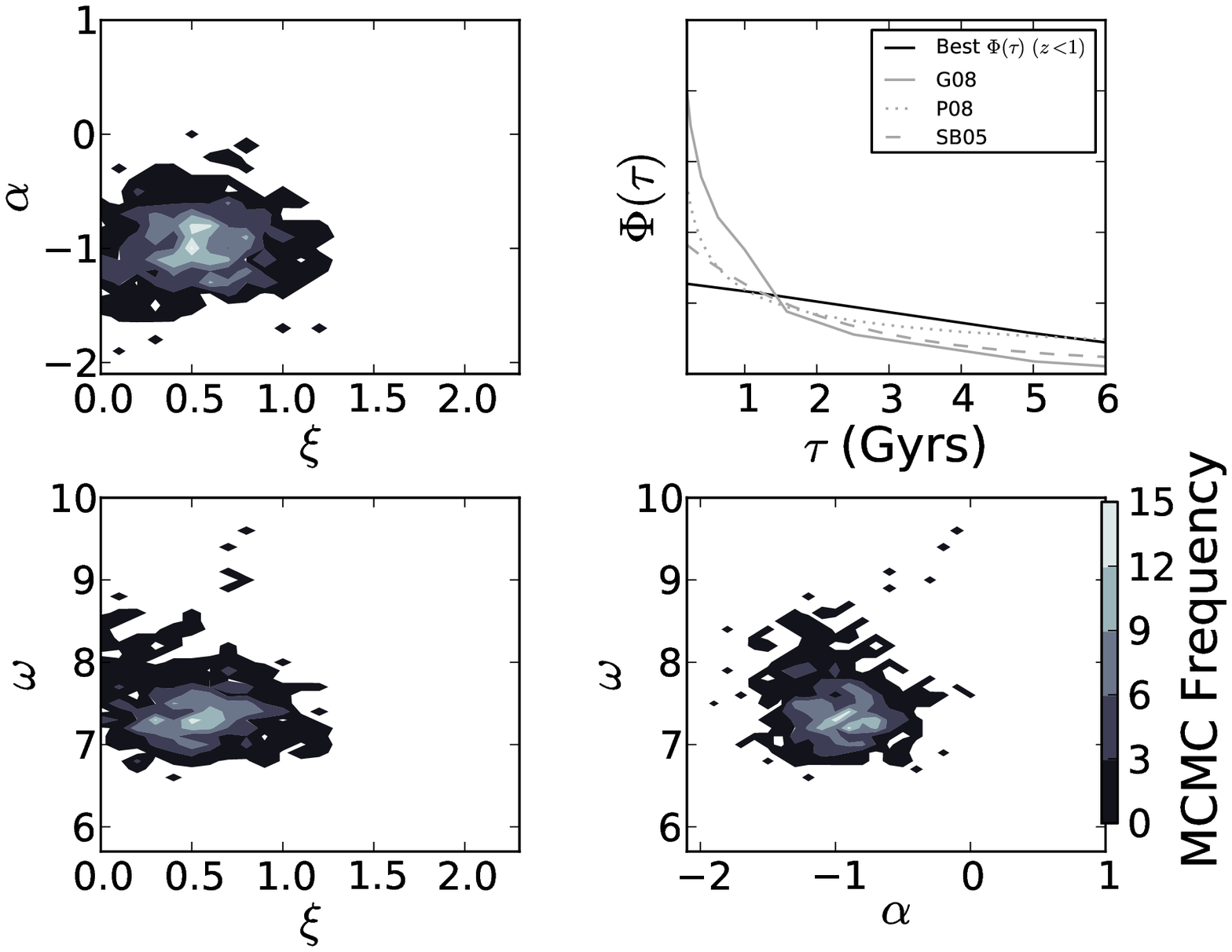}
\caption{\footnotesize Same as for Figure~\ref{contour1} but on the sub-sample of HST SNe~Ia with $z<1.0$.}\label{contour2}
\end{figure*}

\begin{figure*}[t]
   	\centering
   	\includegraphics[width=6in]{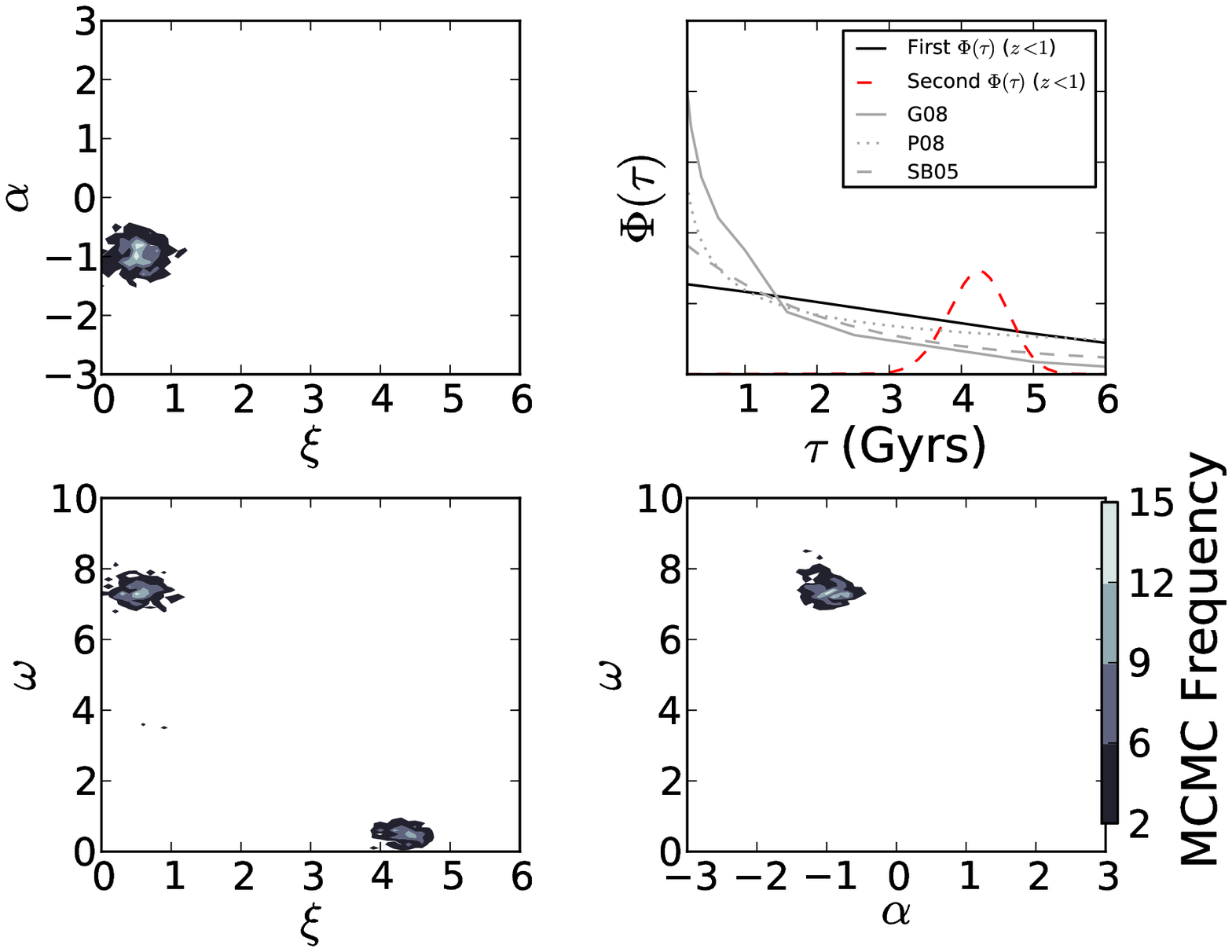}
\caption{\footnotesize Same as for Figure~\ref{contour2}. The axes have been expanded to reveal the second likelihood peak.}\label{contour3}
\end{figure*}

We performed our MCMC test on a sub-sample of the HST-SN survey SNe~Ia, selecting only the 32 events with $z<1.0$. The resulting MCMC likelihood contours  are shown in Figures~\ref{contour2} and~\ref{contour3}. The results show the highest likelihoods for power-law $\Phi(\tau)$ models which show long tails to large delay times (peak at $\xi=0.6$, $\omega=7.5$, $\alpha=-1.0$). These models bare some resemblance in shape to several SD and DD models, including~\citet{2000ApJ...528..108Y},~\citet{2001ApJ...558..351M},~\citet{2005A&A...441.1055G}, and~\citet{2008MNRAS.388..829G}, and the~\citet{2008ApJ...683L..25P} WD availability model. The match is not perfect as literature models are typically steeper,  requiring disproportionately more prompt ($<1$Gyr) SNe~Ia than the results of our MCMC test. The mismatch may be partly due to some rigidity in skew-normal model of our MCMC test, in that the $\Phi(\xi,\omega,\alpha)$ model may not be flexible enough to reproduce broken power-law-like distributions. However, tests of the flexibility of the model (exemplified in Figure~\ref{dtdex}) suggest this is not the case. 

A recent investigation by \cite{2009ApJ...707...74R} comparing the locations of SN~Ia with the distribution of light in $z<0.07$ spiral galaxies (testing the parent population of stars, as is done in~\citealt{2006Natur.441..463F} and~\citealt{2010arXiv1001.5042S} for Gamma-ray Bursts) offers an alternative explanation. Their results seem to indicate a substantial delay for even the potentially prompt SN Ia population of at least 200-500 Myr. Although this is not as large as the delay from the MCMC results shown in~\S\ref{results}, a scenario in which events delayed by $\la0.5-1$ Gyr simply do not occur (essentially cutting all delays below 1 Gyr in the upper-right panel of Figure~\ref{contour2}) would improve the agreement of our $z<1$ sample results with many $\Phi(\tau)$ models.

It is also interesting to note that the $z<1.0$ MCMC results show a second slightly smaller likelihood peak near $\bar{\tau}=4.2$ Gyr ($\xi=4.5$, $\omega=0.5$, $\alpha=-1.0$; see Figure~\ref{contour3}).  It is tempting to interpret the existence of two likelihood peaks as support for bimodality in the intrinsic delay-time distribution, implying two separate SN Ia mechanisms. However the more cautious interpretation is that this MCMC test is not designed to find bimodal $\Phi(\tau)$ models. Its design is to find the peak and distribution for a model that has a single mode, assuming the intrinsic $\Phi(\tau)$ can be accurately characterized in this way. Secondly, the result also only materializes when a sub-sample of $z<1$ data is used, and not with the full dataset. There is currently no reason to suspect that our SNe~Ia at $z<1$ carry more weight than those at $z>1$, although the possibility is discussed further in \S\ref{sensitivity}. The MCMC result from the full sample should be principally considered.

\subsection{Cut on SN~Ia Peak Luminosity}
Another intriguing cut on our sample is to see if the redshift distribution is dependent on event luminosity. The Supernova Legacy Survey noted an interesting trend in which their lowest redshift sample ($0.0<z<0.1$) were dominated by lower luminosity (narrower lightcurve width) events, and their highest redshift sample ($0.75<z<1.5$) by more luminous events~\citep{2007ApJ...667L..37H}. If low-luminsoity events have a substantial fraction of SN~1991bg-like events, and these events represent a separate channel for SN~Ia production (perhaps more delayed than for normal SNe~Ia), then this subset could show a different delay-time distribution than the parent population, reflected in the observed distribution of redshifts for these events. The converse could also be expected for the high-luminosity sample, which should contain a high fraction of SN~1991T-like events that could also represent a separate SN~Ia mechanism. A caveat is that by nature less luminous events will be less numerous in the highest redshift regimes, as magnitude-limited surveys lose sensitivity to them at slightly lower-$z$ (our survey sensitivity is discussed further in \S\ref{sensitivity}).

\begin{figure*}[t]
   	\centering
   	\includegraphics[width=6.0in]{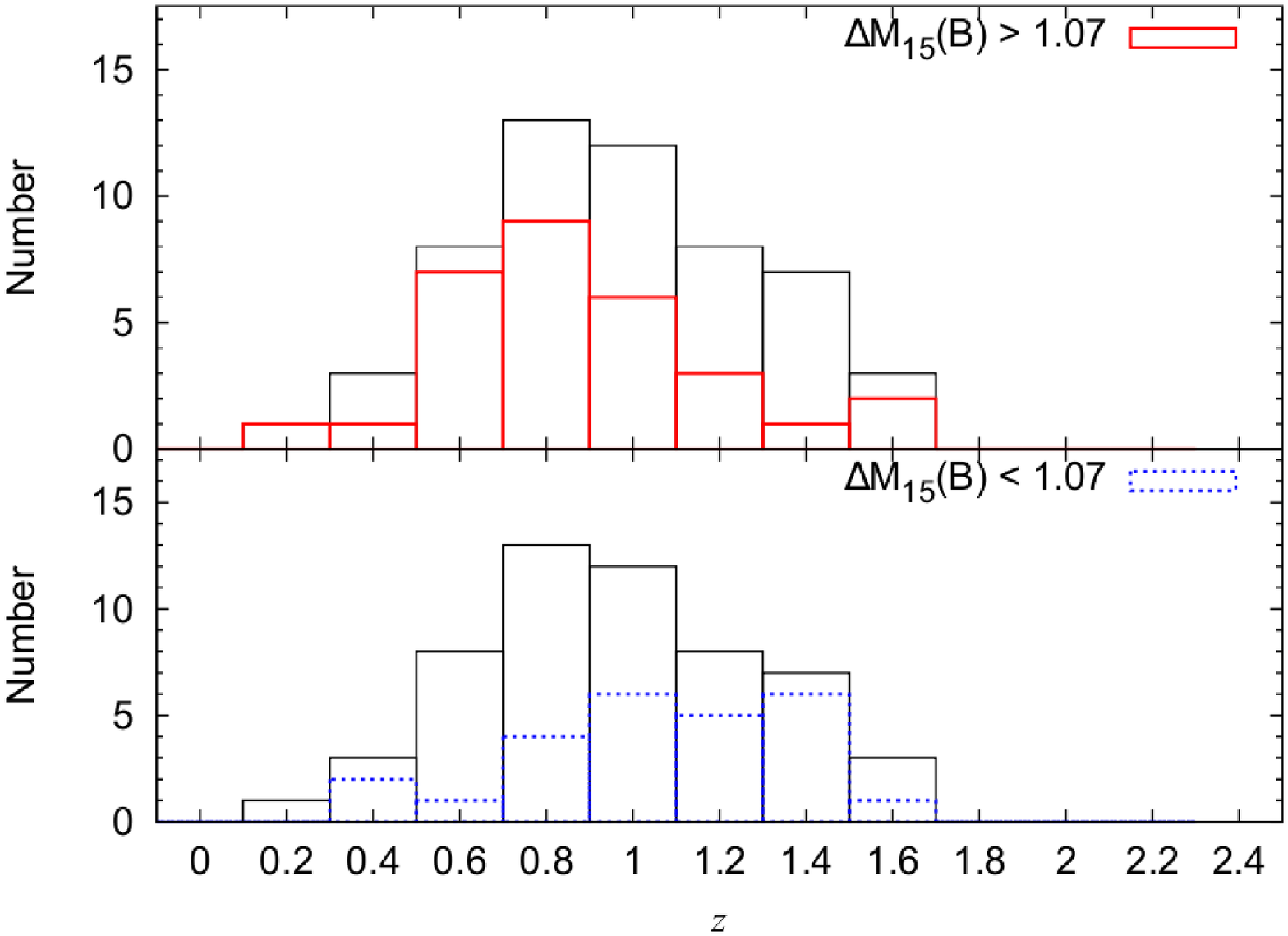}
	\caption{\footnotesize Redshift distribution of SNe~Ia from the HST-SN survey (black histogram). The red histogram shows the distribution for $\Delta {\rm M}_{15}(B)>1.07$, and the blue histogram is for $\Delta {\rm M}_{15}(B)<1.07$, in the rest-frame of each supernova.}\label{red2}
\end{figure*}

Using fits to the lightcurve widths, determined in the rest-frame of each SN, we separate the sample into high-luminsoity (with $\Delta {\rm M}_{15}(B)<1.07$) and low-luminosity ($\Delta {\rm M}_{15}(B) > 1.07$) events.\footnote{$\Delta {\rm M}_{15}(B)=1.07$ is the mode of SN~Ia lightcurve widths~\citep{1999AJ....118.1766P}.} It should be noted that with the exception of the 22 events used for the cosmological investigation~\citep{2007ApJ...659...98R}, most of the remaining supernovae have limited lightcurve information, which limits further precise determination of peak luminosity or decline rates. Figure~\ref{red2} shows the redshift distribution for the high and low luminosity samples, compared to the full sample. As expected, there is a trend similar to~\citet{2007ApJ...667L..37H} for the less luminous supernovae to shift to lower redshifts, peaking near $z\sim0.8$ rather than $z\sim1.0$, and an opposite trend (although less pronounced) in the more luminous sample. These shifts, however, do not translate to significant changes in the derived delay-time distribution function.

\begin{figure*}[t]
   	\includegraphics[width=3.33in]{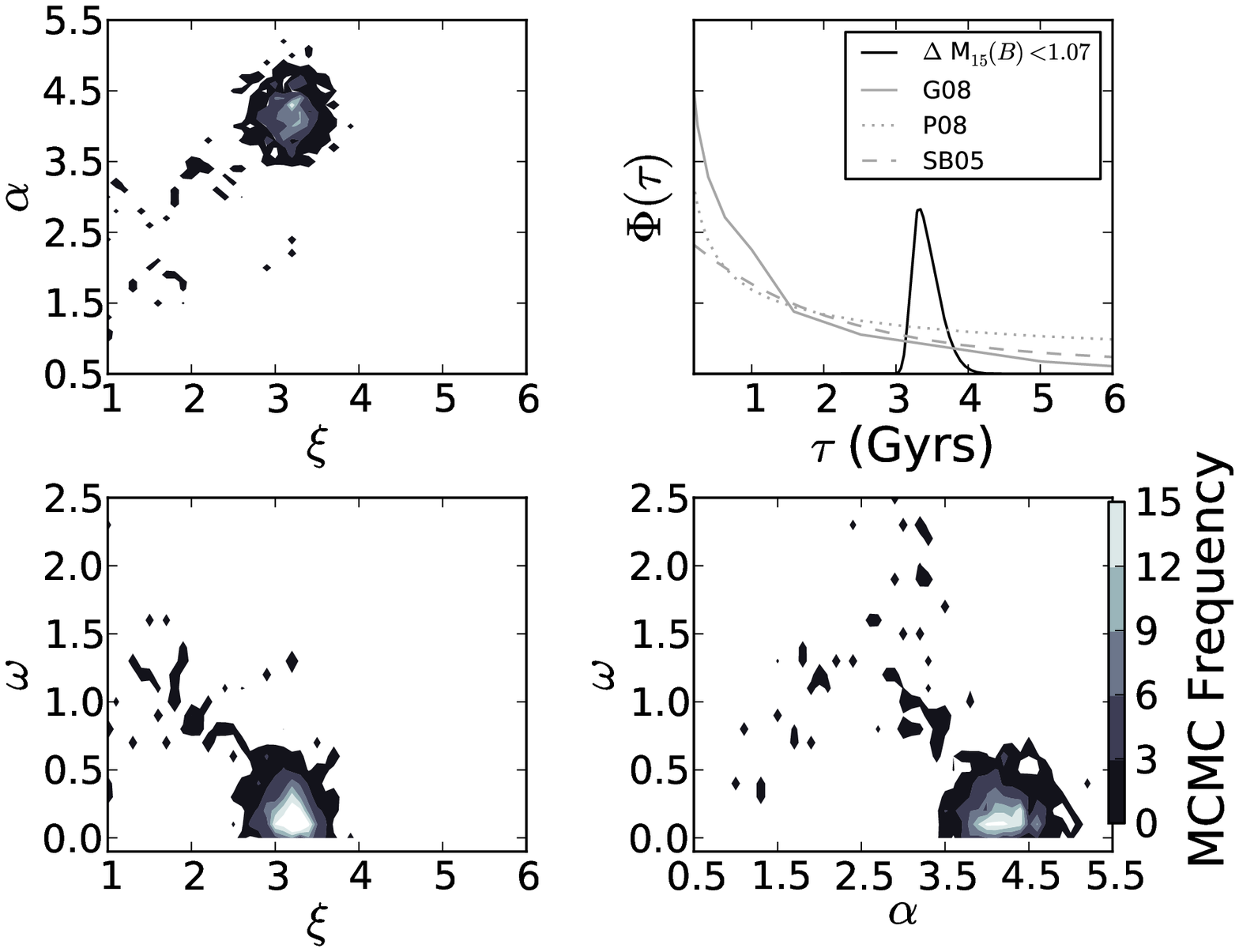}
	\includegraphics[width=3.33in]{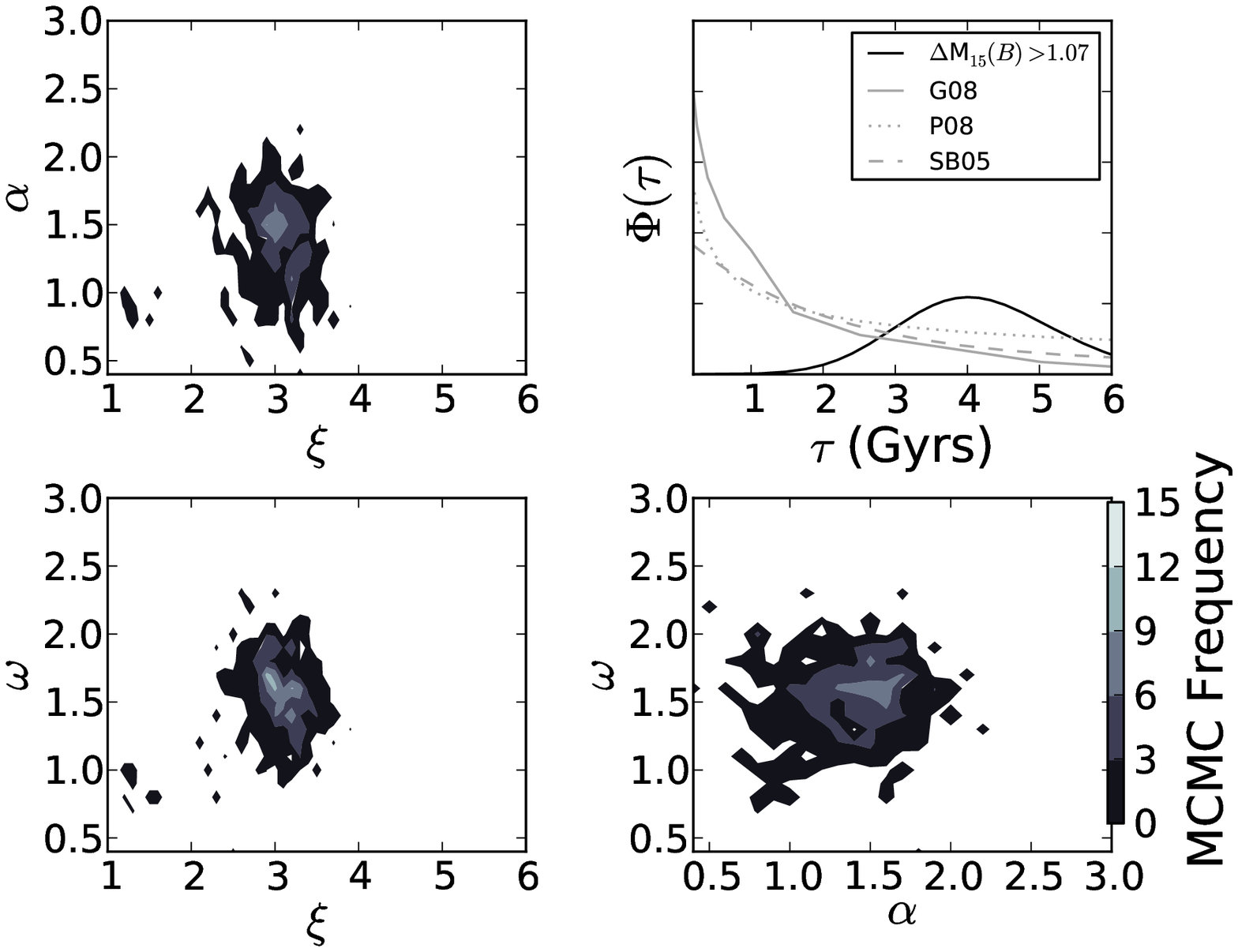}
\caption{\footnotesize Same as in Figure~\ref{contour1}, but for the more luminous $\Delta {\rm M}_{15}(B)<1.07$ sub-sample (left four panels), and the less luminous $\Delta {\rm M}_{15}(B)>1.07$ sub-sample (right four panels).}\label{redcontour}
\end{figure*}

We performed our  MCMC test independently on the bright ($\Delta {\rm M}_{15}(B)<1.07$) sample of 26 events, and the faint ($\Delta {\rm M}_{15}(B)>1.07$) sample of 30 events, the results of which are shown in Figure~\ref{redcontour}. As is shown, the more luminous events do prefer slightly shorter delay times (with a mode near 3.2 Gyr) than the less luminous sample (with a mode near 4.2 Gyr and a broader distribution). The lack of great change is expected, due to appropriate corrections in the control time calculations which truncate the assumed luminosity function\footnote{A gaussian distribution consistent with recent results from the Sloan Digital Sky Survey~\citep{2009arXiv0905.4125Y}.} at high or low luminosity.

\subsection{Sensitivity of the HST-SN Survey}\label{sensitivity}
Arguably, most of the rationale for such large delay times comes from the notable dearth of SNe~Ia $z>1.4$ from the survey. There has been some concern on the completeness and sensitivity of the HST-SN survey to events in the highest magnitude (and redshift) ranges, but recent attempts find more SNe~Ia at $z>1.4$ in {\em the same HST-SN data} via independent detection criteria~\citep{2008ApJ...673..981K}, novel pixel-by-pixel $N(N-1)/2$ comparisons~\citep{2007AAS...211.4725R}, and a thorough review of the deepest region of the GOODS South (the UDF; \citealt{2006AJ....131.1629S}) have failed to produce convincing additional candidates. Additionally, there is further evidence for a lack of $z>1.4$ SNe~Ia from the Subaru Deep Survey~\citep{2007MNRAS.382.1169P}. 

Another  critical concern discussed in~\citet{2008MNRAS.388..829G} is that high redshift galaxies are likely to have more internal extinction than their low-$z$ counterparts due to the very enhanced dust production associated with the high rate of star formation. At these redshifts, we probe the rest-frame UV region of the SN~Ia spectrum, a region which is inherently photon-deficient, and extremely sensitive to host extinction. Naturally, attempts are  made to correct for the portion of SNe~Ia lost to internal extinction  through the control times (essentially ``efficiency corrections''), by using modeled and observed radial distributions of SNe~Ia in low-$z$ galaxies and extinction distributions within them~\citep{1998ApJ...502..177H,1999ApJS..125...73J}. However this concern is for an additional extinction over that which is traditionally accounted for. To date, neither the HST-SN cosmology data~\citep{2007ApJ...659...98R} nor programs which span intermediate redshift ranges (e.g., SNLS and ESSENCE) and bridge the gap from low-$z$ to high-$z$, show any evidence for an ``extinction excess'' trend with redshift. There is, in fact, an opposite tendency for routines such as MLCS2k2 (which fit $R_V$ as a free parameter in lightcurve fitting) to prefer ``grayer'' extinction laws for higher redshift SNe~Ia, with $1.6 < \langle R_V \rangle < 2.7$ \citep{2007ApJ...659..122J}.

\subsection{Changes in the Star-Formation Rate Density, \\ or a Possible Metallicity Effect on SN Ia Production} 
Our best-fit $\Phi(\tau)$ model is very different than what may have been expected from a more classical modeling of the SN~Ia progenitors, and from recent interpretations of lower-$z$ data. It is difficult to see how both could be correct. However, there is a way out of this apparent quagmire. If indeed the standard SD and DD $\Phi(\tau)$ models are correct, then there must be an additional factor which puts additional constraint on the shape of the derived $\Phi(\tau)$ at low $\tau$ in this study.

One very likely possibility is that the shape of the $\dot{\rho}_{\star}(z)$ function (shown in Figure~\ref{fig:sfr}) does not flatten beyond $z>6$, rather declines more sharply.  Very recent (and tentative) results from the HST+WFC3/IR early release portion of an 192-orbit ultra-deep survey of the HUDF (Illingworth, GO11563) seem to indicate that $\dot{\rho}_{\star}(z)$ declines to nearly $z=0$ values near $z=8$~\citep{2010ApJ...709L.133B, 2010ApJ...709L..16O}. Based on test done in~\cite{2004ApJ...613..200S} on alternative models for the $\dot{\rho}_{\star}(z)$ function (including a model with a sharper decline), we predict that similar MCMC test will show greater likelihood values in smaller $\tau$ regions, but the best-fit overall will remain high.

 Another potential culprit could be an innate metallicity effect, which suppresses the prompt component of the natural $\Phi(\tau)$ through some requirement of a ``minimum metallically'' for the SN~Ia mechanism. A physical rationale for a minimum metallicity effect in SD scenarios could be that potential WDs progenitors must develop a sufficient counter-wind to allow for steady mass accretion, to bulk up the core of the WD without triggering surface H \& He flashes (novae) and substantial mass losses, or accretion induced core-collapse supernovae, or other scenarios which ultimately fail to produce a SN~Ia~\citep{1998ApJ...503L.155K, 2008arXiv0801.0215K}.  This is somewhat supportive of the resent observational results of~\citet{2009arXiv0901.4338C} on the large-scale environmental impacts on SN~Ia production, but acts in the opposite way than what would be expected from their results.
 
If metallicity is a critical missing factor, a new parameterization of the SN~Ia rate with redshift could then be:
\begin{equation}
	{\rm R}_{Ia}(t)=\varepsilon\,\int_{t_0}^t \Phi(t-t')\,{\rm \dot{\rho}_{\star}}(t')\,\zeta\{{\rm [O/H]}(t-t')\}\,dt',
\end{equation}

Where $\zeta\{{\rm [O/H]}(\tau)\}$ describes the efficiency in successfully making SNe based on the [O/H] in the system at the time of formation. The [O/H] ratio is chosen as it is suspected that products of the CNO-cycle, specifically $^{22}$Ne which is traced by $^{16}$O abundances in the interstellar medium, have a larger impact on the SN~Ia outcome than Fe-peak elements in the ISM~\citep{2003ApJ...590L..83T}, although it is not at all clear how ISM metallicity impacts SN~Ia luminosity or production. The global metallicity enrichment of the universe should be a slowly decreasing function with lookback time~\citep{2005ApJ...618...68K}, but due to its convolution with the intrinsic delay-time function of SNe~Ia, could provide a steep cutoff to the SN~Ia production at highest-$z$~\citep{2006ApJ...648..884R}.

\section{Conclusions}

The HST-SN data are most supportive of a single dominant mechanism for the production of SN~Ia, which requires between 3 and 4 Gyr of incubation from system formation to explosion. This mechanism is supportive of single degenerate models in which the companion donor stars are low mass ($\la2$ M$_\odot$), based on the main-sequence lifetimes which dominate (much more so than the accretion times) the incubation period.  The data are largely inconsistent with progenitor scenarios with short ($<1$ Gyr) development times, and are inconsistent global scenarios where prompt mechanisms make up a substantial fraction of all channels employed by SN~Ia progenitors.

The preference of our delay time model tests for high $\tau$ are largely motivated by the  observed reduction in $z>1$ SNe~Ia. Further investigations of the SN~Ia rate at even higher redshifts, from $1.5<z<3.0$, should elucidate possible metallicity trends (or other effects) from sensitivity issues, and are plausible in large campaigns with {\it HST} with the IR channel of WFC3. They will also be easily assessable in future space-based missions such as the {\it James Webb Space Telescope}. In the meanwhile,  it is also imperative that the SN~Ia rate measures in the $0.1<z<1.0$ range reach consensus to make more rigorous global comparisons to the SN~Ia rate history, and further refine the empirical delay-time distribution function. This is within reach with large-scale surveys such as the Palomar Transient Factory and Pan-STARRS.  Ultimate comparisons will be achievable when the Large Synoptic Survey Telescope, and the NASA/DOE Joint Dark Energy Mission/International Dark Energy Cosmology Survey are realized. 

\acknowledgments We thank an anonymous referee for valuable comments. This work is based on observations with the NASA/ESA {\it Hubble Space Telescope}, obtained at the Space Telescope Science Institute, which is operated by AURA, Inc., under NASA contract NAS 5-26555. These observations are associated with GO programs 9352, 9425, 9583, 9728, 10189, 10339, 10340, and 10802, and AR-10980. Additional financial support for this work was provided by the Kentucky Space Grant Consortium (K07R10), funded by a NASA Training Grant as part of the National Space Grant College and Fellowship Program, and by the Western Kentucky University Research Foundation.

\end{document}